\documentstyle[prl,aps,floats,epsf,color]{revtex}
\begin{document}
\baselineskip=12pt
\def\be{\begin{equation}}
\def\ee{\end{equation}}
\def\bea{\begin{eqnarray}}
\def\eea{\end{eqnarray}}
\def\E{{\rm e}}
\def\bearst{\begin{eqnarray*}}
\def\eearst{\end{eqnarray*}}
\def\peleven{\parbox{11cm}}
\def\peffec{\peight{\bearst\eearst}\hfill\peleven}
\def\pspace{\peight{\bearst\eearst}\hfill}
\def\ptwelve{\parbox{12cm}}
\def\peight{\parbox{8mm}}
\twocolumn
[\hsize\textwidth\columnwidth\hsize\csname@twocolumnfalse\endcsname

\title
{Dynamics of the Markov Time Scale of Seismic Activity May Provide
\\ a Short-Term Alert for Earthquakes }

\author
{M. Reza Rahimi Tabar,$^{1,2}$ Muhammad Sahimi,$^3$ K.
Kaviani,$^4$ M. Allamehzadeh,$^5$ J. Peinke,$^6$ M.
Mokhtari,$^5$\\M. Vesaghi,$^1$ M.D. Niry,$^1$ F. Ghasemi,$^1$ A.
Bahraminasab,$^1$ S. Tabatabai,$^7$ and F. Fayazbakhsh$^1$}

\vskip 1cm

\address
{$^1$Department of Physics, Sharif University of
Technology, P.O. Box 11365-9161, Tehran, Iran \\
$^2$CNRS UMR 6529, Observatoire de la C$\hat o$te d'Azur,
BP 4229, 06304 Nice Cedex 4, France\\
$^3$ Department of Chemical Engineering and Material Science,
University of
Southern California Los Angeles, CA 90089, USA\\
$^4$Department of Physics, Az-zahra University,
P.O.Box 19834, Tehran, Iran \\
$^5$Department of Seismology, The International Institute of
Earthquake
Engineering and Seismology, IIEES, P.O. Box 19531, Tehran, Iran\\
$^6$Carl von Ossietzky University, Institute of Physics, D-26111
Oldendurg, Germany \\
$^7$Institute of Geophysics, University of Tehran, Iran  \\
}
 \maketitle


\begin{abstract}

{\bf We propose a novel method for analyzing precursory seismic
data before an earthquake that treats them as a Markov process and
distinguishes the background noise from real fluctuations due to
an earthquake. A short time (on the order of several hours) before
an earthquake the Markov time scale $t_M$ increases sharply, hence
providing an alarm for an impending earthquake. To distinguish a
false alarm from a reliable one, we compute a second quantity,
$T_1$, based on the concept of extended self-similarity of the
data. $T_1$ also changes strongly before an earthquake occurs. An
alarm is accepted if {\it both} $t_M$ and $T_1$ indicate it {\it
simultaneously}. Calibrating the method with the data for one
region provides a tool for predicting an impending earthquake
within that region. Our analysis of the data for a large number of
earthquakes indicate an essentially zero rate of failure for the
method.}

\end{abstract}
\hspace{.3in}
\newpage
] Earthquakes are complex phenomena.$^1$ Although still subject to
some debate, precursory anomalies, such as changes in the seismic
recordings, and anomalous variations in the chemical,
hydrological, and electromagnetic properties of the area in which
earthquakes occur, usually precede a large earthquake.$^{1,2}$ One
school of thought believes that the anomalies occur within days to
weeks before the main shock, but probably not much earlier,$^3$
and that the spatial precursory patterns develop at short
distances from impending large earthquakes. A second school
believes that the anomalies may occur up to {\it decades} before
large earthquakes, at distances much larger than the length of the
main shock rupture, a concept closely linked to the theory of
critical phenomena$^{1,2}$ which was advocated$^{1,4,5}$ as early
as 1964 with a report$^4$ documenting the existence of long-range
correlations in the precursors. Knopoff {\it et al.}$^6$ reported
recently the existence of long-range spatial correlations in the
increase of medium-range magnitude seismicity prior to large
earthquakes in California.

Pursuing a model of rock rupture and its relation with critical
phenomena and earthquakes,$^7$ a method of analysis was
introduced$^{8,9}$ that, for certain values of its parameters, led
to a power law (typical of critical phenomena) for the system's
time-to-failure. Several groups$^{10}$ proposed percolation$^{11}$
and hierarchical models of damage/rupture prior to an earthquake.
In particular, Sahimi {\it et al.}$^{12}$ proposed a connection
between percolation, the spatial distribution of earthquakes'
hypocenters, and rock's fracture/fault networks. Sornette and
Sammis$^{13}$ developed a theory according to which the power law
that describes the accelerated seismicity close to a large
earthquake is accompanied by log-periodic correction terms,$^{14}$
which were shown$^{15}$ to also exist in the power law that
describes the increase in the energy that rock releases as it
undergoes fracturing. Such ideas were further developed by Huang
{\it et al.},$^{16}$ with empirical evidence provided by Bowman
{\it et al.},$^{17}$ and view a large earthquake as a temporal
singularity in the seismic time series, resulting from the
collective behavior and accumulation of many previous smaller-size
events.$^{18}$ In this picture, as the stress on rock increases,
micro-ruptures develop that redistribute the stress and generate
fluctuations in it. As damage accumulates, the fluctuations become
spatially and temporally correlated, resulting in a larger number
of significantly-stressed large domains. The correlations
accelerate the spatial smoothing of the fluctuations, culminating
in a rupture with a size on the order of the system's size, and
representing its final state in which earthquakes occur.
Numerical$^{19}$ and empirical$^{20}$ evidence for this picture
indicates that, similar to critical phenomena, the correlation
length of the stress-field fluctuations increases significantly
before a large earthquake. Notwithstanding the evidence, proving
or refuting the notion of earthquakes as a critical phenomenon
entails replacing the proxies, used for checking the build-up of
the cooperativity that leads to large earthquakes, by a direct
measure of the dynamic evolution of the stress field.
Unfortunately, such a procedure is far beyond the present
technical abilities.

A theory of earthquakes should predict, (1) {\it when} and (2)
{\it where} they occur in a wide enough region. It should also be
able to (3) distinguish a false alarm from a reliable one. In this
paper, we propose a method for predicting earthquakes which
possesses the three features. The method estimates the Markov time
scale (MTS) $t_M$ of a seismic time series - the time over which
the data can be represented by a Markov process.$^{21-25}$ As the
seismic data evolve with the time, so also does $t_M$. We show
that the time evolutioon of $t_M$ provides an effective alarm a
short time before earthquakes. The method distinguishes abnormal
variations of $t_M$ {\it before} the arrival of the P-waves, hence
providing enough of a warning for triggering a
damage/death-avoiding response {\it prior to} the arrival of the
more damaging S-waves.

The method first checks whether the seismic data follow a Markov
chain and, if so, measures the function MTS $t_M$.$^{24-25}$
Characterization of the statistical properties of fluctuations of
$n$ measured quantities of the stochastic process $x(t)$ requires
evaluation of the joint probability distribution function (PDF)
$P_n(x_1,t_1;\cdots;x_n,t_n)$. If the data are a Markov process,
then, $P_n= \Pi_{i=1}^{n-1}p(x_{i+1},t_{i+1}|x_i,t_i)p(x_1,t_1)$,
where $p(x_{i+1},t_{i+1}|x_i,t_i)$ are conditional probabilities
such that the Chapman-Kolmogorov (CK) equation,
\begin{equation}
p(x_2,t_2|x_1,t_1)=\int dx_3\;p(x_2,t_2|x_3,t_3)p(x_3,t_3|x_1,t_1)
\end{equation}
holds for any $t_3$ in $t_1<t_3<t_2$. The validity of the CK
equation for describing the process is checked by comparing the
directly-evaluated $p(x_2,t_2|x_1,t_1)$ with the those calculated
according to right side of Eq. (1). To determine $t_M$ for the
data we compute for given $x_1$ and $x_2$ the quantity,
$Q=|p(x_2,t_2|x_1,t_1)-\int dx_3p(x_2,t_2|x_3,t_3)
p(x_3,t_3|x_1,t_1)|$, in terms of, for example, $t_3-t_1$. In
practice, we take $t_1=0$ and $t_3=\frac{1}{2}t_2$, and vary
$t_2$. Plotting $Q$ versus $t_3$ produces the position of $t_M$ in
the limit $Q\to 0$.$^{25}$

Our analysis of seismic data (see below) indicates that the
average $t_M$ for the {\it uncorrected} background seismic time
series is much smaller than that for earthquakes data (P-wave plus
S-wave). Thus, at a certain time before an earthquake, $t_M$ rises
significantly and provides an alarm for the earthquake. As we show
below, the alert time $t_a$ is on the order of hours, and depends
on the earthquake's magnitude $M$ and the epicenter's distance $d$
from the data-collecting station(s).

The sharp rise in $t_M$ at the moment of alarm is, in some sense,
similar to the increase in the correlation length $\xi$ of the
stress-field fluctuations in the critical phenomena theories of
earthquake, since $t_M$ is also the time over which the events
leading to an earthquake are correlated. Therefore, just as the
correlation length $\xi$ increases as the catastrophic rupture
develops, so also does $t_M$. However, whereas it is exceedingly
difficult to directly measure $\xi$, $t_M$ is computed rather
readily. Moreover, whereas $\xi$ is defined for the {\it entire}
rupturing system over long times, $t_M$ is computed {\it online}
(in real time), hence reflecting the correlations of the most
recent events that are presumably most relevant to an impending
earthquake.

To distinguish a false alarm that might be indicated by $t_M$ from
a true one, we use a second time-dependent function that we
compute based on the extended self-similarity (ESS) of the seismic
time series.$^{24,25}$ The ESS is characterized by $S_p$, a
structure function of order $p$, defined by
\begin{equation}
S_p(\tau)=\langle|x(t+\tau)-x(t)|^p\rangle\sim\langle|x(t+\tau)-x(t)|^3
\rangle^{\zeta_p}\;
\end{equation}
where $\tau$ is the lag (in units of data points). The first
nontrivial moment (beyond the average and variance) of a
distribution is $S_3$, and because for a Gaussian process,
$\zeta_p=\frac{1}{3}p$, the deviations from this relation
represent non-Gaussian behavior. It is also well-known$^{26,27}$
that the moments $S_p$ with $p<1$ contain information on frequent
events in a time series. Prior to an earthquake the number of
frequent events (development of cracks that join up) suddenly
rises, indicated by a sudden change in $S_p$ with $p<1$. We
observe that the starting point of $S_p(\tau)$ ($p<1$) versus
$S_3(\tau)$ is different for different type of data set.$^{26,27}$
To determine the distance form the origin we define the function
$T_1=T(\tau=1)=[S^2_{0.1}(\tau=1)+S^2_3(\tau =1)]^{1/2}$. Close to
an earthquake the function $T_1(t)$, also estimated online,
suddenly changes and provides a second alert. Its utility is due
to the fact that it is estimated very accurately even with very
few data points, say 50, hence enabling online analysis of the
data collected over intervals of about 1 second. Thus, even with
few data points, the method can detect the change of correlations
in the incoming data. For example, for correlated synthetic data
with a spectral density $1/f^{2\alpha-1}$, one obtains
$T_1=-7\alpha +7$.

%

\begin{figure}
\epsfxsize=8.5truecm\epsfbox{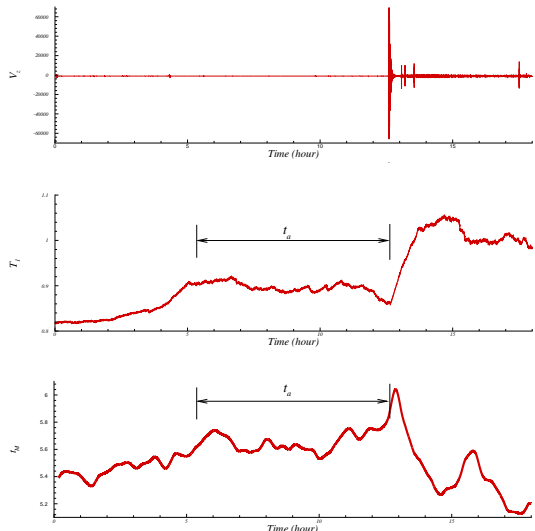} \narrowtext
\caption{Time-dependence of $T_1$ and $t_M$ for a recent
earthquake of magnitude 6.3 in northern Iran, and their comparison
with the vertical ground velocity data $V_z(t)$. $t_M$ is in
number of data points (the frequency at the station is 40 Hz),
$T_1$ is dimensionless, while $V_z(t)$ is in ``counts'' which,
when multiplied by a factor $1.1382\times 10^{-3}$, is converted
to $\mu$m/sec. The sensors were (broad-band) Guralp CMG-3T that
collect data in the east-west, north-south and vertical
directions. The thresholds are $t_{Mc}=5.6$ and $T_{1c}=0.88$. }
 \end{figure}


We have analyzed the data for {\it vertical ground velocity}
$V_z(t)$ for 173 earthquakes with magnitudes $3.2\leq M \leq 6.3$
that occurred in Iran between $28^\circ$N and $40^\circ$N
latitude, and $47^\circ$E and $62.5^\circ$E longitude, between
January 3 and July 26, 2004. Recorded by 14 stations, the data can
be accessed at http://www.iiees.ac.ir/bank/bank$\_$2004.html. The
frequency was 40 Hz for 2 of the stations and 50 Hz for the rest.
The vertical ground velocity data were analyzed because with our
method they provide relatively long (on the order of several
hours), and hence useful, alarms for the impending earthquakes.
Fourty (discrete) data points/second are recorded in the
broad-band seismogram for the vertical ground velocity $x(t)\equiv
V_z$. To analyze such data and provide alarms for the area for
which the data are analyzed, we proceed as follows.

(1) The data are analyzed in order to check whether they follow a
Markov chain [the directly-computed $p(x_2,t_2|x_1,t_1)$ must be
equal to the right side of Eq. (1)]. (2) The MTS $t_M(t)$ of the
data are estimated by calculating the residual $Q$ of the CK
equation (see above). For long-enough data series ($10^3$ data
points or more) the function $t_M(t)$ are estimated where $Q\to
0$, but for shorter series the minimum in $Q$ provides estimates
of $t_M(t)$. (3) $T_1(t)$ is computed for the same data. To
compute $S_p(\tau)$ (we used $p=1/10$) the data $x(t)$ are
normalized by their standard deviation, hence making $T_1$
dimensionless. (4) Steps (1)-(3) are repeated for a large number
of previously-occurred earthquakes of size $M$ at a distance $d$
from the station, referred to as ($M,d$) earthquakes. Earthquakes
with $M<M_c$ and $d>d_c$ are of no practical importance and are
ignored (we used $M_c=4.5$ and $d_c= 150$ km). (5) Define the
thresholds $t_{Mc}$ and $T_{1c}$ such that for $t_M>t_{Mc}$ and
$T_1>T_{1c}$ one has an alert for an earthquake ($M>M_c,d<d_c$).
If $t_{Mc}$ and $T_{1c}$ are too large no alert is obtained,
whereas one may receive useless alerts if they are too small. By
comparing the data for all the earthquakes with $M>M_c$ registered
in a given station, $t_{Mc}$ and $T_{1c}$ for the earthquakes are
estimated. (6) Real-time data analysis is performed to compute the
function $t_M(t)$ and $T_1(t)$. An alarm is turned on if
$t_M>t_{Mc}$ and $T_1>T_{1c}$ {\it simultaneously}. When the alarm
is turned on, it indicates that an earthquake of magnitude $M\geq
M_c$ at a distance $d\leq d_c$ is going to occur. The procedure
can be carried out for {\it any} station. The larger the amount of
data, the more precise the alarm will be.

Figure 1 presents $T_1(t)$ and $t_M(t)$ for an $M=6.3$ earthquake,
occurred on May 28, 2004 at 12:36 am in Baladeh at (36.37N,
51.64E, depth 28) in northern Iran. The data were collected at
Karaj station (near Tehran, Iran) at a distance of 74 km from the
epicenter, and a depth of 70 m. The earthquake catalogue in the
internet address given above indicates that, for several days
before the main event, there was no foreshock in that region.
Thus, $T_1$ and $t_M$ provided a seven hour alarm for the Baladeh
earthquake. Since the data used for computing $t_M$ and $T_1$
were, respectively, in strings of 200 and 50 points, there is no
effect of the events {\it before} they were collected and, hence,
the patterns in Fig. 1 reflect the events taking place in the time
period in which the data were collected.

To estimate the alert times $t_a$, which are on the order of
hours, we carried out an analysis of online data for 14 stations
in Iran's broad-band network (the sensors are Guralp CMG-3T
broad-band), analyzing the vertical ground velocity data. Our
analysis indicates that $t_a$ depends on $M$, being small for low
$M$, but quite large for large $M$. Using extensive data for the
Iranian earthquakes with $M\geq 4.5$ and $d\leq$ 150 km, we have
obtained an approximate relation for the broad-band stations,
shown in Figure 2 and represented by
\begin{equation}
\log t_a=-1.35+2.4\log M\;,
\end{equation}
where $t_a$ is in hours. The numerical coefficients of Eq. (3) for
each area should be estimated from the data collected for that
area. The above analysis can clearly be extended to all the
stations around the world. This is currently underway for Iran's
network. For an earthquake of magnitude $M=4.5$, Eq. (3) predicts
an alert time of about 2 hours. Thus, if, for example, three hours
after the alarm is turned on, the earthquake has not still
happened, we know that the magnitude of the coming earthquake is
$M\geq 5.7$.

In summary, we have proposed a new method for analyzing seismic
data and making predictions for when an earthquake may occur with
a magnitude $M\geq M_c$ at a distance $d\leq d_c$. The method is
based on computing the Markov time scale $t_M$, and a quantity
$T_1$ calculated based on the concept of extended self-similarity
of the data, and monitoring them {\it online} as they evolve with
the time. If the two quantities exceed their respective critical
thresholds $t_{cM}$ and $T_{c1}$, estimated based on analyzing the
data for the previously-occurred earthquakes, an alarm is tuned
on. We are currently utilizing this method for Iran's stations. To
do so, we calibrate the method with the data for the stations in
one region (i.e., estimate $t_{cM}$ and $T_{c1}$ for distances
$d<d_c$). If in a given region there is a single station, then
once the online-computed $t_M$ and $T_1$ exceed their critical
values, the alarm is turned on. If there are several stations,
then once they declare that their $t_M$ and $T_1$ have exceeded
their thresholds, the alarm is turned on. If after about 2 hours,
no earthquake has occurred yet, then we know that the magnitude of
the incoming earthquake will be greater $M_c=4.5$ at a distance
$d<d_c$.

Over the past two years, the method has been utilized in the
Iranian stations. Our analysis indicates that the method's failure
rate decreases to essentially {\it zero} when $t_M$ and $T_1$
provide {\it simultaneous} alarms. That is, practically every
earthquake that we have considered, including those that have been
occurring while we have been performing online analysis of their
incoming data and providing alarms for them (with $M>M_c$), was
preceded by an alarm. Of all the earthquakes that we have analyzed
so far, the method has failed in only {\it two} cases. In our
experience, if after 10 hours no earthquake occurs, we count that
as a failed case. However, as mentioned, we have so far had only
two of such cases.

Finally, it must be pointed out that the most accurate alarms are
obtained from stations that receive data from depths of $>$ 50 m,
and are perpendicular to the active faults that cause the
earthquake, since they receive much more correlated data for the
development of the cracks than any other station.

We are particulary grateful to K.R. Sreenivasan, R. Mansouri, S.
Sohrabpour and W. Nahm for useful discussions, comments, and
encouragement. We would also like to thank M. Akbari, F. Ardalan,
H. Arfaei, J. Davoudi, R. Friedrich,  M. Ghafori-Ashtiany, M. R.
Ghaytanchi, K. Hesami, N. Kamalian, V. Karimipour, A. Mahdavi,
Amalio F. Pacheco, M. Rezapour, A. Sadid Khoy, F. Shahbazi, J.
Samimi, H.R. Siahkoohi, N. Taghavinia, and M. Tatar for useful
comments.

%
 \begin{figure}
  \epsfxsize=8
  truecm\epsfbox{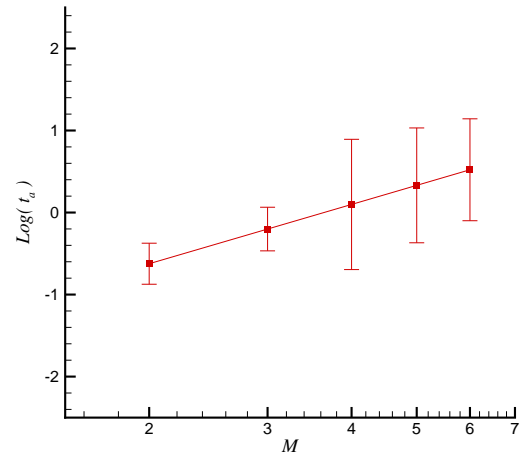}
    \narrowtext \caption{ The dependence of alert time $t_a$ (in hours)
on the magnitude $M$ of the earthquakes, obtained based on the
data from {\it broad-band} stations by analyzing 173 earthquakes
with magnitudes $3.2\leq M \leq 6.3$ that occurred in Iran between
$28^\circ$N and $40^\circ$N latitude, and $47^\circ$E and
$62.5^\circ$E longitude, between January 3 and July 26, 2004. }
  \end{figure}

\end{document}